\documentclass[fleqn,usenatbib,useAMS]{mnras}

\usepackage{etex}
\usepackage[usenames,dvipsnames]{xcolor}
\usepackage{graphicx}	
\usepackage{amsmath}	
\usepackage{amssymb}	
\usepackage{multicol}        
\usepackage{multirow}
\usepackage{graphicx}
\usepackage{bm}		
\usepackage{pdflscape}	
\usepackage[T1]{fontenc}
\usepackage{tgbonum}
\usepackage{float}
\usepackage{siunitx}
\usepackage{appendix}

\graphicspath{{../figures/}}

\definecolor{Red}{rgb}{0.65,0.08,0.05}
\definecolor{Green}{rgb}{0.15,0.45,0.25}
\definecolor{Pink}{rgb}{1.0,0.05,0.5}
\definecolor{Purple}{rgb}{0.3,0.,0.5}

\usepackage[T1]{fontenc}
\usepackage{ae,aecompl}

\usepackage{newtxtext,newtxmath}

\title[Co-evolution of filaments and halos]{The impact of evolving cosmic filaments on mass and spin evolution of dark matter halos}

\author[Jhee et al.]{Hannah Jhee$^{1}$,
Hyunmi Song$^{2}$\thanks{E-mail: hmsong@cnu.ac.kr},
Clotilde Laigle$^{3}$\thanks{E-mail: laigle@iap.fr},
Christophe Pichon$^{3,4}$,
Corentin Cadiou$^{3}$,
Ena Choi$^{1}$
\\
$^{1}$Department of Physics, University of Seoul, 163 Seoulsiripdaero, Dongdaemun-gu, Seoul 02504, Republic of Korea\\
$^{2}$Department of Astronomy and Space Science, Chungnam National University, 99 Daehak-ro, Yuseong-gu, Daejeon 34134, Republic of Korea \\
$^{3}$Institut d'Astrophysique de Paris, 98 bis Boulevard Arago, F-75014 Paris, France \\ 
$^{4}$Kyung Hee University, Department
of Astronomy \& Space Science, Yongin-shi, Gyeonggi-do 17104, Republic of Korea \\  
}

\date{Last updated 2025 May 21; in original form 2025 May 21}

\pubyear{2025}

\begin{document}
\label{firstpage}
\pagerange{\pageref{firstpage}--\pageref{lastpage}}
\maketitle

\begin{abstract}
The evolution of galaxies is closely tied to that of their host dark matter halos, which is in turn strongly modulated by the surrounding large-scale environment. Cosmic filaments are expected to influence the peculiar motions, mass assembly and angular momentum of nearby halos through highly anisotropic matter flows. In order to fully capture the dynamic interplay between the filaments and halos, we develop an algorithm to trace the progenitors of individual filaments identified at $z=0$ with {\tt DisPerSE} in a cosmological $N$-body simulation, by quantifying the spatial similarity between a descendant filament and progenitor candidates. This enables us to reconstruct filament-by-filament evolutionary histories, including their bulk drift and the evolution of radial density profiles, from which splashback radii ($\sim 1\,\rm cMpc$) and core overdensities ($1+\delta\sim 20$) are derived. Using these time-dependent properties, we re-examine halo phase-space trajectories in a filament-centric frame that evolves with time. This eliminates biases inherent to static models by separating halo motions from the bulk motion of the filaments, allowing trajectories to be identified more reliably. We find that as halos approach high-density filaments, their mass accretion rates are systematically suppressed beginning at the filament outskirts, suggestive of early tidal stripping or suppressed net accretion. Furthermore, the evolution of halo spin alignments exhibits a clear departure from stochastic random-walk expectations. This suggests that distinct mass flow regimes in and around filaments exert different torques on infalling halos, thereby changing their angular momentum.
Our findings, derived from a sample screened for major mergers, highlight the `pure' dynamical impact of the filamentary environment. Ultimately, we demonstrate that tracking the simultaneous co-evolution of filaments and halos is essential for accurately characterizing environmental effects.
\end{abstract}


\section{Introduction} \label{sec:intro}
Growing from primordial density fluctuations \citep{Klypin1983, Bond1996}, the cosmic web is structured into several distinct components: clusters (or nodes), which have collapsed gravitationally along all three dimensions; filaments and walls, which have undergone partial collapse in two and one dimension(s), respectively; and voids, which are underdense regions where matter is predominantly expanding rather than collapsing.
These environments span a wide range of densities and tidal conditions, thereby shaping the formation and subsequent evolution of halos and galaxies embedded within them.
A growing body of work indicates that halo and galaxy growth is regulated not only by the amplitude of the (angularly averaged) local density, but also by the anisotropy of the surrounding matter distribution---particularly within filaments and sheets \citep{Hahn2007, Pichon2011, Wang2011, Borzyszkowski2017, Paranjape2018}.
Although filaments occupy only a small fraction of the cosmic volume ($\sim 5$\%), they contain more than 50\% of the total mass \citep{Cautun2014}.
Recent studies further suggest that filaments contribute dominantly to the gravitational and tidal fields over much of the cosmic volume, including substantial fractions of void and sheet regions \citep{Platen2008, Feldbrugge2023, Kugel2024}, extending their dynamical influence well beyond their immediate vicinity.

One of the most noticeable manifestations of filament influence is the tendency for halo and galaxy properties to correlate with the geometry of nearby filaments.
In particular, halo and galaxy spins are known to exhibit mass-dependent alignments with respect to filaments, with lower-mass halos ($\log (M_{\rm halo}/M_{\odot})\lesssim 12.3$) preferentially aligning their spin with the closest filament, while more massive ones exhibit anti-alignment \citep[e.g.][]{Aragon-Calvo2007, Sousbie2008, Codis2012, Laigle2015, GaneshaiahVeena2018, Lopez2020, Welker2020}.
Beyond angular momentum, filamentary environments have also been shown to regulate star formation activity and stellar mass growth, indicating that galaxy evolution depends on cosmic web environments beyond halo mass alone.
Both observations and simulations suggest that galaxies located near filaments experience systematically different star formation rates and quenching efficiencies compared to galaxies in more isotropic environments at fixed halo mass and local density, reflecting the role of filaments as anisotropic channels of gas supply and environmental regulation \citep[e.g.,][]{Malavasi2017, Kraljic2018, Laigle2018, Song2021}.
Recently, \citet{Storck2024} argued that halo spin and shape orientations respond more sensitively to filament environments than halo mass or virial radius alone.
Furthermore, \citet{Tudorache2025}, using the MIGHTEE survey, reported that galaxy spins are more strongly aligned with nearby filaments than statistically predicted by cosmological simulations, and presented evidence that galaxies can orbit around filaments.
Together, these results highlight that filaments are not merely passive overdense regions but dynamically active structures that influence halo and galaxy evolution through sustained anisotropic accretion, tidal interactions and environmental regulation.

Despite these advances, the majority of previous studies have relied on correlations measured at fixed snapshots, implicitly interpreting spatial trends with filament distance as evolutionary sequences.
While such approaches are effective in establishing statistical associations, they provide limited insight into how halos and galaxy properties are built over time in relation to filaments, and which physical processes are responsible for these correlations.
In particular, fixed-redshift-based analyses cannot distinguish whether measured trends reflect the cumulative impact of prolonged interaction with filaments or arise from transient environmental effects.
As a result, the physical pathways through which filamentary environments influence halo and galaxy evolution remain poorly constrained.

In \citet[][J22 hereafter]{Jhee2022}, we moved beyond snapshot-based analyses by explicitly tracking individual halos in time, enabling a time-dependent view of halo-filament interactions.
This Lagrangian approach allowed us to identify distinct dynamical pathways, distinguishing halos that become gravitationally bound to filaments from those that are not bound.
We showed that halos, while they orbit around filaments, could undergo systematic mass loss during pericentric passages due to tidal stripping, and that filamentary environments can suppress halo mass growth, consistent with earlier measurements of filamentary shear effects \citep[e.g.][]{Hahn2009, Borzyszkowski2017}.
We further demonstrated that the observed mass segregation of halos towards filament spines \citep[e.g.][]{Malavasi2017} can be strengthened naturally from differences in formation times and subsequent dynamical evolution, rather than requiring it to be imprinted ab initio.

A key simplifying assumption in J22 was that filaments were treated as static structures, with their positions fixed to those identified at $z=0$.
In reality, filaments can drift and deform in response to large-scale flows driven by surrounding structures.
Anchoring the reference frame to present-day filaments can therefore bias the inferred filament-centric distances and halo trajectories, especially when tracing halos back to high redshift over long lookback times.
On short timescales, a halo's motion near a filament is set primarily by the local filament gravitational potential, so filament drift is a secondary effect for describing instantaneous orbits.
However, over long timescales, ignoring filament evolution can bias measurements of the initial separation and the time of pericentric passage.

In addition, stacking halo trajectories without accounting for the diversity and evolution of filament properties (e.g., thickness or density) may mix signals from physically different filament environments.
Because filaments are continuously evolving structures—for example, their characteristic splashback radii evolve with redshift \citep[e.g.,][]{Ramsoy2021, Wang2024}—we therefore adopt a consistent Lagrangian framework to assess halos’ host filaments more unambiguously.

A necessary first step towards this goal is to develop a methodology to identify the progenitors of individual filaments and to follow their evolution across redshift.
An initial attempt in this direction was presented by \citet{Galarraga-Espinosa2024}.
They identified $z=0$ filaments from the galaxy distribution using \texttt{DisPerSE} \citep[Discrete Persistent Structure Extractor;][]{Sousbie2011} and traced their progenitors by following the evolution of the critical points (CPs) that define the filament network, i.e., locations where the density gradient vanishes ($\nabla\rho=0$).
This approach enabled pioneering studies of filament length evolution and the redshift dependence of radial density profiles.
However, because filaments are defined from the galaxy distribution and traced using a skeleton constructed at a fixed persistence threshold, establishing smooth and robust progenitor-descendant connections can become increasingly challenging at high redshift, where the halo population is sparse and filamentary structures are expected to be less persistent.
These limitations motivate the development of improved filament evolution tracking methods that can more consistently capture continuous geometric and dynamical evolution of individual filaments over cosmic time.

In this paper, we develop a filament evolution tracking framework that identifies and follows the progenitors of individual filaments across cosmic time.
Using this framework, we revisit halo evolution in a reference frame defined by evolving filaments, enabling a more physically consistent interpretation of halo-filament interactions and their cumulative impact on halo growth.
In Section \ref{sec:2}, we introduce our algorithm to identify and trace the progenitors of individual filaments defined at $z=0$.
Using these filament progenitors, we quantify the drift velocity and the evolution of density profiles of selected sets of filament progenitors in Section \ref{sec:3}.
We then revisit in Section~\ref{sec:4.1} the halo motions in a dynamically evolving filament-centric frame, and directly compare the inferred trajectories to those under the fixed-filament assumption of J22.
The mass evolution of halos under the effects of the filament environment is also discussed.
In Section \ref{sec:4.2}, we extend our analysis to the evolution of halo spin alignments with their host filaments.
We summarize our main findings and discuss their implications in Section \ref{sec:5}.

\section{Methods} \label{sec:2}
\subsection{Extracting filaments} \label{subsec:2.1}
This work relies on a dark-matter-only cosmological simulation, $N$-Cluster Run\footnote{\href{https://data.kasi.re.kr/vo/N_cluster_run/}{https://data.kasi.re.kr/vo/N\_cluster\_run/}}.
The size of the simulation box is $120\,h^{-1}\,\rm cMpc$ and the mass of a dark matter particle is $1.072\times 10^{9}\,h^{-1}\,\rm M_{\odot}$.
Halos are identified using {\tt Rockstar} \citep{Behroozi2013a}, and their merger trees are constructed using {\tt consistent-trees} \citep{Behroozi2013b}.
The details of the simulation can be found in \citet{Chun2022}.
Instead of utilizing the whole simulation box, we select three regions around the most massive cluster-like halos, whose masses are approximately $\sim 4.5\times 10^{14}\,h^{-1}\,\rm M_{\odot}$ in order to focus on the prominent filaments around the clusters.
The side length of each selected box is set to be $20R_{\rm vir}$ ($\sim 31\,h^{-1}\,\rm cMpc$).

\begin{figure*}
    \includegraphics[width=0.7\linewidth]{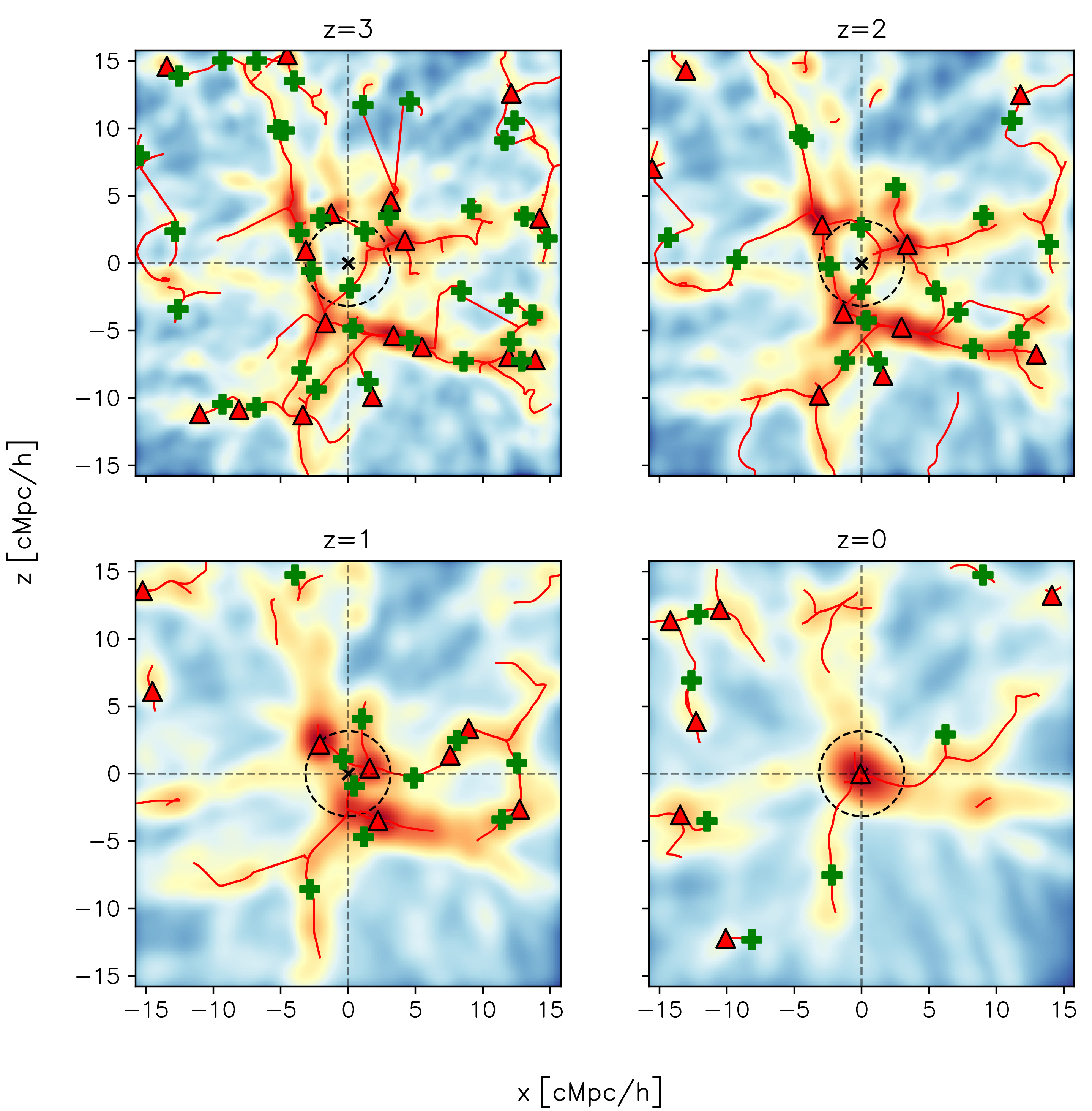}
    \centering
    \caption{The smoothed density significance map of a slab of $\Delta y\sim 4.2\,h^{-1}\rm\,Mpc$ at $z=0,\,1,\,2$ and $3$, around the most massive cluster. Peaks and saddle critical points within the slab are shown as red triangles and green crosses respectively, and filament segments also within the slab are shown as red lines. Here, the {\tt -cut} parameter to filter low-contrast filaments was set to 1. The black crosses and black dashed lines represent the position and $2R_{\rm vir}$ region of the cluster, at $z=0$. Note that in this Figure, the filaments can be attached to nothing if the critical points exist outside of the slab.}
    \label{fig:filament_extraction}
\end{figure*}

We apply {\tt DisPerSE} \citep{Sousbie2011} to extract filament structures.
A filament is defined as a persistent topological structure connecting a density maximum to a saddle point.
Instead of constructing a scale-free density estimator using delaunay tessellation, we calculate a fixed-scale density grid, $\rho(z,\mathbf{x})$, with uniform pixel size of $0.1\,h^{-1}\,\rm cMpc$.
The density field is smoothed using the Gaussian kernel $W(\mathbf{x};R)$ with a smoothing length $R_{\rm sm}(z)$:
\begin{equation}
    \rho_{\rm sm}(z,\mathbf{x})=\int d^3\mathbf{x}'\,\rho(z,\mathbf{x}')W(|\mathbf{x}-\mathbf{x}'|;\,R_{\rm sm}(z)).
\end{equation}
Here $R_{\rm sm}(z)$ is selected so that it effectively smooths out the structures smaller than the minimum size of group-scale halos at each redshift.
Filaments below this scale are not necessarily spurious, but we decided to eliminate those filaments in our sample because they tend to be traced poorly due to the complex histories and low density contrast.
By fixing the minimal scale at which filaments are extracted, we are restricting our filament samples to the most prominent ones connecting groups and clusters.
This also facilitates tracing these prominent structures smoothly over time.
To achieve this, we measure the fraction of halos with $M>10^{13}h^{-1}\,\rm M_{\odot}$ (group/cluster-scale halos) at $z=0$, which is $\sim 0.2\%$.
At each redshift, the minimum virial radius, $R_{\rm group, min}(z)$, of the top $0.2\%$ massive halos is used to calibrate the smoothing scale as $R_{\rm sm}(z)=1.5R_{\rm group,min}(z)$.
Here, the factor $1.5$ is used as a proxy to convert the virial radius into the splashback radius of a halo \citep{More2015}.
At $z=3$, $2$, $1$ and $0$, the smoothing scales correspond to $0.66$, $0.64$, $0.52$ and $0.43 h^{-1}\,\rm cMpc$, respectively.
We then transform the smoothed density field into a log-significance field $\eta(z,\mathbf{x})$ as below:
\begin{equation}
    \eta(z,\mathbf{x})=\frac{\log\rho_{\rm sm}(z,\mathbf{x}) - \langle\log\rho_{\rm sm}\rangle(z)}{\sigma_{\log\rho_{\rm sm}}(z)},
\end{equation}
where $\langle\log\rho\rangle$ is the mean of the logarithmic density field, and $\sigma_{\log\rho}$ is its standard deviation.
This normalization ensures that the input fields deviate only slightly from a Gaussian distribution with zero mean and unit variance.
By normalizing the density field in each snapshot as such, filaments extracted at higher redshifts with the same persistence threshold as at $z=0$ correspond to less prominent ones but are statistically equivalent to the $z=0$ filaments in terms of persistence.
By setting the persistence cut threshold ({\tt -cut}) to 1.0, we identify respectively 167, 230 and 213 filaments at $z=0$ around each of the three most massive clusters.
By construction, we can always find a counterpart departing from the saddle point ({\tt type=2} CP) for a given filament, constructing a filament that connects two nodes ({\tt type=3} CP) through the saddle point.
To avoid edge effects, filaments near the boundaries of the extracted sub-volumes around clusters are removed from our dataset, with their critical points lying outside the volume.
As a result, we are left with 223 node-to-node filaments at $z=0$ around the three most massive clusters in the simulation.
The filaments extracted at $z=3$, $2$, $1$ and $0$ are shown in Figure \ref{fig:filament_extraction}, on top of the smoothed density fields.

\subsection{Tracing filaments through cosmic time}
In this section, we briefly summarize how we traced the progenitors of filaments through cosmic time.
The details of the algorithm are described in the Appendix \ref{sec:ap:A}.

We first developed a metric to quantify the spatial similarity between two one-dimensional structures embedded in three-dimensional space.
This metric is based on evaluating the degree of overlap between the spatial extents, or equivalently the spatial distributions, of the two structures along each of the three coordinate axes.
Specifically, for each axis, we divide the coordinate range into bins and compute the fraction of the total filament segments in each bin, for a descendant and a progenitor candidate.
We then define a score by multiplying the two binned segment fractions and summing the resulting products over all bins, such that a value of 0 indicates no overlap and a value of 1 indicates complete overlap.
Denoting the scores along the three axes by $S_x$, $S_y$, and $S_z$, respectively, we define the overall score as $S=(S_xS_yS_z)^{1/3}$, so that it becomes large only when the overlap is high along all three axes.
Using this scoring system, we can evaluate the spatial similarity between a given filament at $z=0$ and the filaments identified in the immediately preceding snapshot by tracing the snapshots backward in time.
In this way, we select the progenitor with the highest probability of being the ancestor of the target filament, namely the one with the highest score.

By applying this scoring scheme consecutively across a series of snapshots, one can ideally construct the most plausible progenitor sequence for a given filament.
In practice, however, the filament length can depend on the choice of the persistence threshold used for filament extraction.
In other words, if the nodes connected to the two ends of a filament are not identified at a given threshold, the filament length may change discontinuously relative to that in the preceding snapshot.
When no progenitor was found at the adopted threshold, or when the length of the identified progenitor exhibited an anomalous discontinuity, we lowered the persistence threshold in a given snapshot and repeated the progenitor search including filaments with lower density contrast. 
Among the 223 filaments present at $z=0$, 72 filaments could be traced back to $z>1$.
For the remaining filaments, the lack of a traced progenitor beyond $z=1$ may reflect either that they formed during the later stages of cosmic evolution ($z<1$) or that the tracing algorithm was unable to identify their progenitors reliably.

\begin{figure*}
    \includegraphics[width=\linewidth]{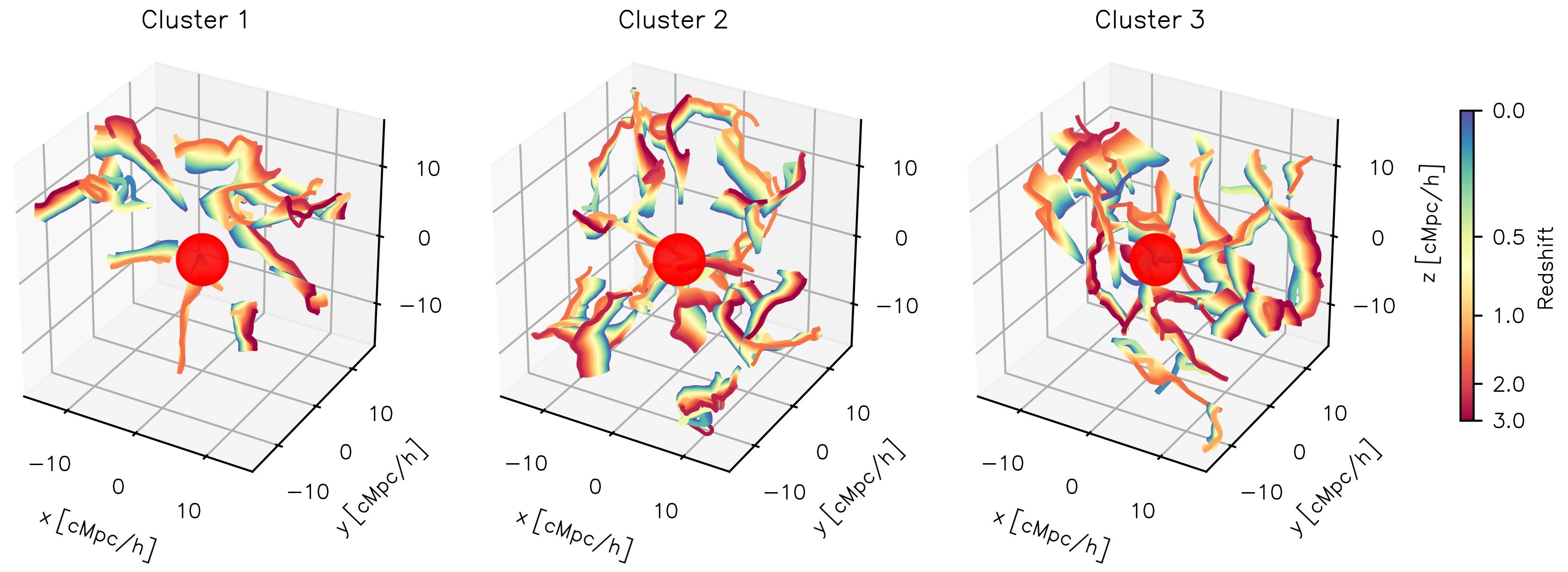}
    \caption{Spatial distribution of the 72 filaments at $z=0$ whose progenitors have been traced back to $z>1$. The red sphere at the center denotes a radius of $2R_{\rm vir}$ from each cluster center at $z=0$; the three clusters have virial radii of 1.58, 1.57, and 1.56 $\mathrm{cMpc}/h$, respectively. Filaments at $z=0$ are shown in blue, while filaments at higher redshift are shown as progressively red.}
    \label{fig:5_filaments_spatial}
\end{figure*}

The spatial distribution of these filaments and their progenitors at higher redshifts is shown in Figure \ref{fig:5_filaments_spatial}, where the color encodes redshift with red corresponding to $z=3$ and blue to $z=0$ (as indicated by the colorbar).
In the following, we examine how the drift motion and density profiles of filaments evolve over time, highlighting a small number of representative cases for clarity while retaining the full filament sample for statistical analysis.

\section{The evolution of filaments}\label{sec:3}

\subsection{Drifting filaments}\label{sec:3.1}

\begin{figure}
    \includegraphics[width=\linewidth]{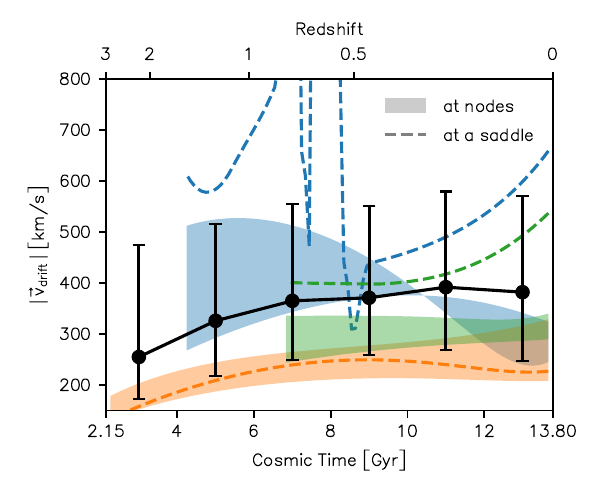}
    \caption{The peculiar drift velocities of the three filaments around the third most massive cluster, shown as a function of cosmic time. The velocities are measured at three critical points: two nodes (shaded regions) and one saddle point (dashed lines). The shaded regions between the two curves illustrate the ranges of velocity variation between the two nodes, while the dotted lines denote the velocities measured from the saddle point. The black error bars represent the median and the 16th  and 84th percentiles of the full sample of 72 filaments.}
    \label{fig:drift_velocity}
\end{figure}

While filaments form through matter infall along their two principal axes, the residual transverse motion associated with the infall can lead to the overall drift of filaments \citep{Pichon2011}.
Measuring a filament's drift velocity\footnote{Throughout this paper, all velocities refer to peculiar velocities. Writing the physical position as $\mathbf{x}_{\rm phys}=a\mathbf{x}_{\rm com}$, the physical velocity is $$\mathbf{v}_{\rm phys}=\dot{\mathbf{x}}_{\rm phys}=\dot{a}\mathbf{x}_{\rm com} + a\dot{\mathbf{x}}_{\rm com}$$ where the first term corresponds to the Hubble flow and the second term is the peculiar velocity.} --as distinct from the underlying matter flow-- is intrinsically non-trivial because a filament cannot be treated as a rigid body.
As it evolves, a filament may stretch, contract or bend \citep{Galarraga-Espinosa2024}, such that different parts of the structure do not share a single coherent velocity.
In this context, tracking the displacement of an individual node alone can lead to biased velocity estimates: changes in the filament's length or orientation may introduce apparent velocity components that are unrelated to the global drift, and in some cases may even appear perpendicular to it.
For the same reason, defining the drift velocity from the average position of filament segments can also be misleading, as this average is sensitive to elongation or shortening of the structure rather than to translational motion.
Tracking the displacement of a saddle point presents additional challenges.
Although saddle points are topologically well-defined, their positions are more susceptible to local density fluctuations and substructure than those of nodes, leading to less stable temporal tracking.

Given these limitations, rather than assigning a single drift velocity to each filament, we present the velocity measurements of the two nodes and the saddle point separately.
Figure \ref{fig:drift_velocity} illustrates three representative filaments, shown in different colors, while the full sample of 72 filaments is displayed as black error bars representing the median and 16th-84th percentiles.
In Figure \ref{fig:drift_velocity}, the velocity differences between the two nodes are indicated by shaded regions, while the velocities inferred from the saddle point are shown with dotted lines.
The saddle-based velocities are noticeably less stable than those derived from the nodes, demonstrating that, unlike nodes--which effectively correspond to group/cluster-like halos--saddle points are more strongly influenced by their surrounding environments.
Although node-based estimates are generally more robust, some filaments still exhibit significant scatter, which motivates our choice to highlight only representative cases rather than displaying all individual measurements.

The measured drift velocities in our sample span roughly $200-600\, \rm km\,s^{-1}$, with noticeable variations among individual filaments.
These magnitudes are broadly consistent with the expectation that large-scale transverse motions in the cosmic web can give rise to residual pattern drift at scales of filaments \citep{Pichon2011, Shandarin2011}.

Examining the temporal evolution of individual filaments, we find no universal monotonic trend in the drift velocity.
Among the three simple cases, some filaments exhibit increasing drift velocities, others show a decline, while some display an initial increase followed by a plateau.
The diversity reflects the fact that filament drift is governed by local and time-dependent gravitational environments, including the growth rates of connected nodes and interactions with surrounding structures.

In contrast, when considering the full sample of 72 filaments, the median distribution reveals a clear broadening of the drift velocity range with cosmic time.
Rather than indicating a uniform acceleration of filament drift, this trend suggests an increasing heterogeneity in the dynamical states of filaments as structure formation proceeds.
As the cosmic web evolves into a more interconnected and asymmetric network, individual filaments experience increasingly distinct dynamical influences, leading to a wider dispersion of drift velocities.

A potential caveat is sample-selection bias. 
If filaments in denser environments experience more frequent interactions and mergers—owing to a higher abundance of massive halos (or nodes)—then their spines may be more difficult to trace robustly to high redshift. 
In that case, the subset of filaments that can be safely tracked beyond $z\sim 1$ may be biased toward lower-density environments, and our drift-velocity estimates could represent a lower bound on the full filament population. 
Quantifying the magnitude of this bias will require a larger filament sample and/or alternative progenitor-tracking schemes, which we defer to future work.

\subsection{Evolution of density profiles} \label{sec:3.2}
The evolution of the density profiles of large-scale filaments has been investigated in \citet{Galarraga-Espinosa2024} and \citet{Wang2024}, who examined the redshift evolution of stacked radial density profiles of filaments.
In this work, we instead focus on the evolution of the radial density profiles of individual filaments.

We use dark matter particles, instead of halos, in order to calculate the radial density profiles $\rho(d_{\rm fil})$.
Filaments are modeled as a series of small cylindrical segments rather than a single straight cylinder.
As halos lying within filaments produce local density peaks that generate small-scale noise and bias the filament density estimate, particles belonging to halos ($<r_{\rm vir}$) are excluded from the total mass in each cylindrical bin.
The volume occupied by the removed particles is likewise subtracted from the volume of each cylindrical bin, considering the position with respect to the boundaries of each cylinder.
Excluding the halo component from a cylindrical shell is challenging because a halo may extend beyond the width of the shell.
To handle this, we first computed the cumulative radial density profile, $\bar{\rho}(<d_{\rm fil})$, and then obtained the differential density, $\rho(d_{\rm fil})$, by differentiating the cumulative profile under the assumption of cylindrical symmetry:
\begin{equation}
    \rho(d_{\rm fil})=\bar{\rho}+\frac{d_{\rm fil}}{2}\frac{d\bar{\rho}(<d_{\rm fil})}{d(d_{\rm fil})}.
\end{equation}
We also smoothed the density profiles by fitting them with 6th-order polynomials, as done in \citet{Wang2024}.

\begin{figure*}
    \centering
    \includegraphics[width=0.99\linewidth]{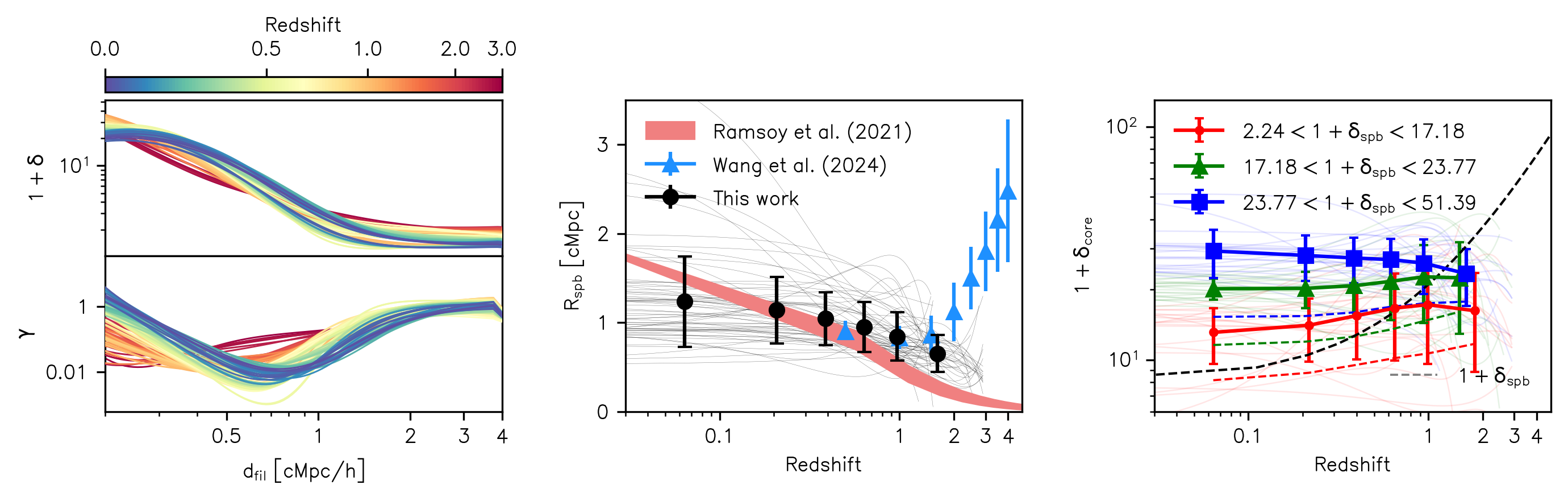}
    \caption{\textit{Left}: The overdensity profiles at all redshifts of one filament are shown in the top panel, and the slope $\gamma=d\log\rho/d\log d_{\rm fil}$ in the bottom panel. In each panel, the red curve corresponds to $z=3$ and the blue one  to $z=0$. The splashback radius is defined as the radius minimizing $\gamma$. \textit{Middle}: The evolution of the splashback radii of 72 filaments are shown as thin gray lines. The average is shown as black circles with $1\sigma$ errors, while the binning is done such that the number of data points be the same. Assuming a scaling of the form $\propto (1+z)^{\alpha}$, the best-fit power-law slope for the evolution of radii is $\alpha=-0.70\pm 0.45$. As a comparison, the pink shaded region shows the filament core radius evolution from \citet{Ramsoy2021}: $r_0\propto (1+z)^{-2.18\pm0.28}$ (in comoving) fitted from $z\sim 8$ to $z\sim 3.5$. Here, the formula has been extrapolated to $z\sim 0$, and the normalization has been fixed to match our data at $z\sim 0.2$. The blue triangles are the results from \citet{Wang2024}. \textit{Right}: The evolution of the core density within the fixed radius $0.2h^{-1}\,\rm cMpc$ for each filament is shown as thin solid lines in the background. The filaments are subsampled according to their overdensity at $z=0$, and their average evolutions are tracked separately in thick solid lines with the corresponding $1\sigma$ errors in red (low density), green (mid-density) and blue (high density). The density and redshift bins are chosen such that each bin contains the same number of data points. For comparison, the evolutions of the density within the splashback radius is shown as dashed lines with the same colors. The slope could be compared to the black dashed line, which is the evolution of the splashback density solely due to the evolution of splashback radius (therefore, $\propto (1+z)^{1.4}$).}
    \label{fig:radii_evol_denprof}
\end{figure*}

The upper left panel of Figure~\ref{fig:radii_evol_denprof} shows the evolution of the radial density profile of a filament (red for $z=3$ and blue for $z=0$).
The corresponding logarithmic slope, $\gamma=d\log\rho/d\log  d_{\rm fil}$, is shown in the lower part of the leftmost panel.
The splashback radius ($R_{\rm spb}$), defined as the radius at which $\gamma$ reaches its minimum \citep{Diemer2014, Diemer2015, More2015}, is presented in the middle panel, in comparison with the results of \citet{Wang2024} and the redshift scaling reported in \citet{Ramsoy2021}.
While \citet{Wang2024} covers a slightly higher redshift range than in this work, the inferred evolutionary trend is broadly consistent across the overlapping redshift range.
We note that the rapid increase in filament widths at high redshift reported by \citet{Wang2024} is not observed for most of the filaments in our sample.
This discrepancy may arise from differences in how the density profiles are constructed.
In \citet{Wang2024}, density profiles are measured by stacking across the full filament population, whereas we compute profiles for individual filaments.
The latter approach can be more susceptible to fluctuations induced by nearby substructures, because azimuthal averaging at fixed radius can be strongly affected by localized overdensities.

This effect is illustrated in Figure \ref{fig:cross_section}, which shows the dark-matter particle distribution in cross-sections of a representative filament at several redshifts, together with the corresponding splashback radius.
At $z\sim 3$, the filament environment contains more prominent substructures than at $z\lesssim 1$.
When the matter distributions are stacked over many filaments, such substructures can contribute to excess density at larger radii and thereby bias the inferred splashback radius high, while simultaneously smoothing out filament-to-filament variations.
This may help explain why the increase of width at high redshift seen in \citet{Wang2024} is weaker or absent in our individually measured profiles.

\citet{Ramsoy2021} extended the analysis to higher redshift and found an empirical scaling for the dark matter filament core radius, $r_0\propto (1+z)^{-3.18\pm 0.28}$, in physical units, where $r_0$ is defined via the Plummer-like radial profile $\rho(r)\propto [1+(r/r_0)^2]^{-2}$.
As a simple extrapolation, we extend this relation to our redshift range and find the predicted slope to be consistent with our measurements.
In our sample, the splashback radius (quoted in comoving units) increases from $\sim 0.5\,\rm cMpc$ at $z\sim 2$ to $\sim 1\,\rm cMpc$ by $z=0$.
The redshift dependence of our data points (black error bars) is best fitted by a power-law scaling of the form $(1+z)^\alpha$, with $\alpha=-0.70\pm 0.45$.
\citet{Ramsoy2021} inferred the filament core radius by fitting an analytic model to individual cross-section profiles, and they followed the evolution of a single `main' filament feeding the galaxy rather than stacking over a filament population.
This methodological similarity to our analysis may also contribute to the weaker high-redshift widening reported in their study, compared to population-stacked measurements.
Although the average radius evolution is broadly consistent with the stacked results, the substantial deviations exhibited by several individual filaments show that filament evolution can vary considerably, likely reflecting differences in their surrounding environments.

Observational studies have also reported redshift evolution in filament widths, although the exact trend depends on detailed methods.
\citet{Choi2010}, using 97 filaments identified in the DEEP2 survey at $z\sim 0.8$ and 958 filaments from SDSS at $z\sim 0.1$, found that the width distribution broadens and shifts toward smaller widths with cosmic time.
This trend is opposite to the mild increase of $R_{\rm spb}$ found in our work.
However, a direct comparison is not straightforward, because \citet{Choi2010} measured widths from the galaxy distribution using a Hessian-based filament finder, whereas our $R_{\rm spb}$ is defined from the minimum logarithmic slope of dark matter particle density profiles around individual progenitors.
The broadening of the observed width distribution is nevertheless consistent with the large filament-to-filament diversity seen in our sample.

More recently, a similar tendency toward smaller radii at lower redshift has also been reported by \citet{Aguerri2026}, who measured galaxy overdensity profiles around SDSS filaments and found smaller mean radii at $0.05<z<0.1$ than at $0.1<z<0.3$.
These observational results therefore highlight that filament-size evolution can depend sensitively on the tracer population, the filament-finding method, and the definition of `width'.
Therefore, our splashback-based radius should be regarded as a dynamical boundary of the dark-matter filament environment rather than as a direct analogue of observational galaxy-defined filament widths.

In the rightmost panel of Figure~\ref{fig:radii_evol_denprof}, we present the evolution of the filament core density, $1+\delta_{\rm core}$, measured within a fixed comoving radius $R_{\rm core}=0.2\,h^{-1} \,\rm cMpc$.
The filaments are divided into three groups according to their $z=0$ core density: the highest-density group is shown in blue, the intermediate-density group in green, and the lowest-density group in red.
We find that the highest-density filaments undergo a progressive concentration toward the core, whereas the lowest-density filaments exhibit a decline in core mass.
In contrast, the core density of intermediate-density filaments remains nearly unchanged over time.
We compare this behavior to the splashback density contrast, $1+\delta_{\rm spb}$, defined as the density enclosed within the splashback radius.
For the same three filament groups, $1+\delta_{\rm spb}$ is shown by dashed curves in the corresponding colors.
Additionally, because the splashback radius itself evolves with redshift, we overplot as a black dashed line the expected trend if the enclosed mass is held fixed and only the splashback radius evolves.
Using the fitted redshift scaling ($\alpha=-0.7$), this implies $R^{-2}\propto(1+z)^{1.4}$, which serves as a reference for comparison with density changes arising purely from radial expansion.
At $z\gtrsim 3$, the decline in $1+\delta_{\rm spb}$ is substantially slower than would be expected from the radius-only evolution scenario.
This behavior is consistent with continued mass supply from the surrounding walls into filaments being effective at high redshift, partially compensating for the dilution associated with the increasing filament radius.
By contrast, at $z\lesssim 0.3$, the intermediate- and low-density filaments show density decreases at a rate comparable to the radius-only expectation, suggesting that net accretion from the surroundings has significantly weakened.
Meanwhile, the highest density filaments continue to show a slower decline than the other two groups, indicating that they remain dynamically active and may still be supported by ongoing accretion and/or deeper gravitational confinement.

\section{Analyzing Halo Evolution} \label{sec:4}
In J22, the progenitors of the filamentary structures were not identified, and halo motions were therefore analyzed relative to filaments that were fixed at their $z=0$ positions.
In this section, we revisit halo dynamics by incorporating the newly available progenitor information, allowing halo motions to be analyzed with respect to the evolving filaments at each epoch.
As a result, all properties considered below--such as filament density and halo distance to the filament spine--are defined instantaneously from the corresponding filament progenitors, rather than their present-day counterparts.
We examine the impact of instantaneous filament core density on the mass evolution of nearby halos.
Also, we analyze the alignment of the halo spins and the filament axis.
From the 72 filaments selected in the previous section, we collect 552 halos which approach closer than $2h^{-1}\,\rm cMpc$ at $z=0$, which is the same criterion as in J22.

\subsection{Revisiting J22}\label{sec:4.1}
\subsubsection{Phase-space analysis}\label{sec:4.1.1}
\begin{figure}
    \centering
    \includegraphics[width=\linewidth]{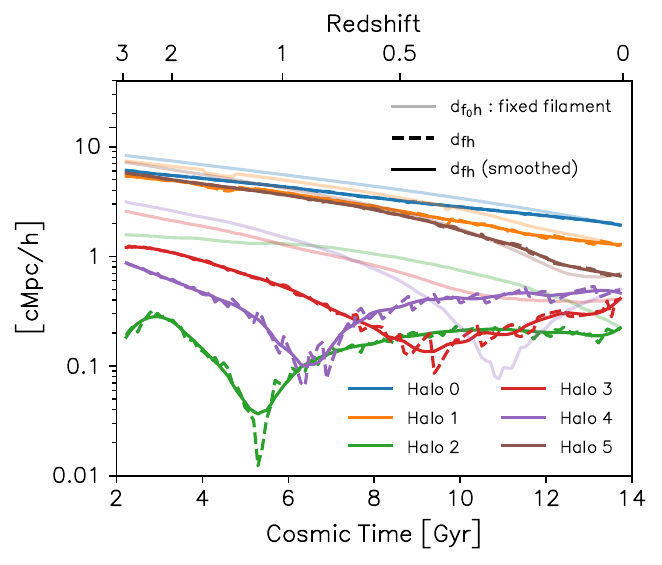}
    \caption{For six halos around one of the 72 filaments, the distance from halos to filaments is shown as a function of cosmic time (bottom label) or redshift (top label). The faint solid lines indicate the distances assuming the filament positions are fixed at $z=0$, while opaque dashed curves are measured using the positions of the filament progenitors. Because the curves are noisy, we also smoothed the trajectories using Savitzky-Golay filter, which performs polynomial fitting over a specified number of consecutive data points in the one-dimensional array, preserving the geometry of the function as much as possible while smoothing the data.}
    \label{fig:t_versus_dfh}
\end{figure}

Whereas J22 measured the filament--halo separation relative to filaments fixed at $z=0$, which we denote as $d_{\rm f_0h}$\footnote{This quantity was denoted as $r_{\rm perp}$ in J22; we use $d_{\rm f_0h}$ here for consistent notation.}, 
in this work we measure the instantaneous separation, $d_{\rm fh}$, to the corresponding progenitor filament identified at each snapshot. 
A direct comparison between the two is shown in Figure~\ref{fig:t_versus_dfh} for six representative halos that lie within $2\,h^{-1}\,{\rm Mpc}$ of one of our $z=0$ filaments.
In the figure, $d_{\rm f_0h}$ from the fixed-filament assumption is shown as faint solid curves (with darker solid curves indicating the smoothed trajectories), while $d_{\rm fh}$ is shown as dashed curves. 
In all six cases, fixing filaments at $z=0$ leads to systematically larger inferred separations, with the discrepancy becoming most pronounced at high redshift, where filament drift relative to the $z=0$ spine is non-negligible. 
This bias directly affects the identification of the pericentric passage, defined here as the time when the filament-centric distance reaches a minimum. 
For halos 2--4, which already reside within $\sim 1\,h^{-1}\,{\rm Mpc}$ of the filament at their first appearance, the inferred pericenter time can be shifted by more than $4\,{\rm Gyr}$ (halos 3 and 4) or missed entirely (halo 2) when using $d_{\rm f_0h}$.
Therefore, while the fixed-filament assumption may still provide a reasonable description of the overall orbital motions of halo populations, 
our results show that it can lead to substantial inaccuracies for individual halos -- particularly those formed in close proximity to the filament. 
Consequently, analyses that aim to recover detailed halo trajectories should avoid relying on a fixed-filament framework.

\begin{figure}
    \centering
    \includegraphics[width=\linewidth]{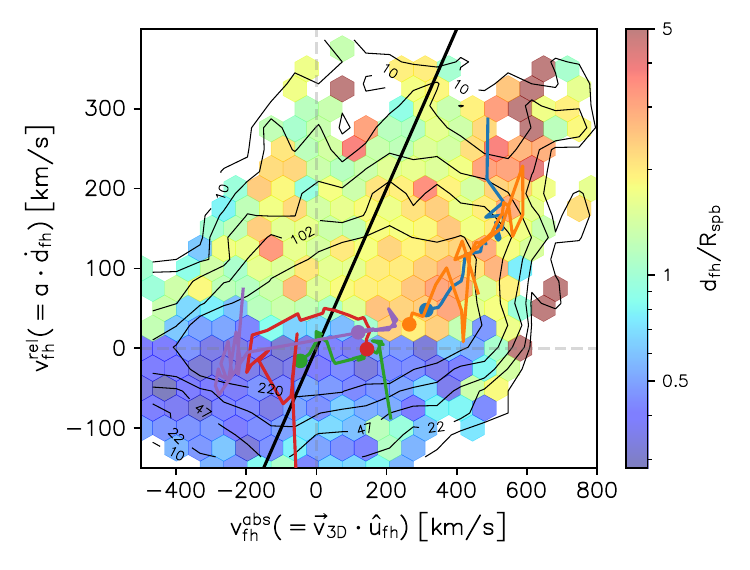}
    \caption{The comparison between the absolute velocity components along the filaments, $v_{\rm fh}^{\rm abs}=\mathbf{v}_{\rm 3D}\cdot\hat{u}_{\rm fh}(z)$, and the relative velocity, $v_{\rm fh}^{\rm rel}=a\cdot d(d_{\rm fh})/dt$. The black diagonal line marks the one-to-one relation. The underlying distribution is shown as black contours, with five levels equally spaced in logarithmic scale between 10 and 220. The colors of hexagonal bins show the median normalized distances. The trajectories of the six halos from Figure \ref{fig:t_versus_dfh} are shown, where initial position in the diagram is marked as circles with the same color. The trajectory represents the median of five snapshots to reduce the noises.}
    \label{fig:vperp_compare}
\end{figure}

In Figure \ref{fig:vperp_compare}, we compare two estimators of the halo velocity along the direction to the nearby filament.
The $x$-axis is obtained by projecting the halos' three-dimensional velocity onto the unit vector $\hat{u}_{\rm fh}$ connecting the halo to the filament, while the $y$-axis is given by the time derivative of the halo-filament separation shown in Figure \ref{fig:t_versus_dfh}, i.e., $v_{\rm fh}^{\rm rel}=a\,({\rm d}d_{\rm fh}/{\rm d}t)$, with $a$ denoting the scale factor.
This corresponds to the relative peculiar radial velocity ``in the filament rest-frame".
In this case, in order to avoid unphysical velocity spikes due to unstable traces of distance, we use smoothed $d_{\rm fh}$ to calculate $v_{\rm fh}^{\rm rel}$.
The hexagonal bins are colored by average $d_{\rm fh}/R_{\rm spb}$ at the positions, and only those with more than 10 data points are shown.
We also showcase the six halos in Figure \ref{fig:t_versus_dfh} using the same colors on top.

We note that the deviation from one-to-one line in Figure \ref{fig:vperp_compare} corresponds to the filaments' drift velocity along this direction.
Recalling that the filament drift velocities measured in Section \ref{sec:3.1} could span from $\sim 200\,\rm km/s$ to $\sim 600\,\rm km/s$, we see that the filament velocity component along $\hat{u}_{\rm fh}$ inferred in Figure \ref{fig:vperp_compare} is of the same order and consistent with those values in the context.
Going further from the analysis in Section \ref{sec:3.1}, the comparison between the values on the $x$- and $y$-axis here also allows us to infer whether a halo is moving broadly in the same direction as its host filament ($v_{\rm fh}^{\rm rel}<v_{\rm fh}^{\rm abs}$) or in the opposite.
By jointly inspecting the color scale of the hexagonal bins and the distribution traced by the contours, we find that halos with $d_{\rm fh}\gtrsim 2-3\,R_{\rm spb}$ lie predominantly below the one-to-one curve, i.e. in the region where $v_{\rm fh}^{\rm rel}<v_{\rm fh}^{\rm abs}$.
This indicates that, even at such large separations (corresponding to earlier cosmic times), halo velocities remain coupled to the local bulk flows along the filaments.
A mild gradient in normalized distance is visible along the $y$-axis in this region, showing that halo velocities increase as they approach the filaments.
Once halos lie within $R_{\rm spb}$ of the filaments, they populate a broad region spanning from $|v_{\rm fh}^{\rm rel}|<|v_{\rm fh}^{\rm abs}|$ to $|v_{\rm fh}^{\rm rel}|>|v_{\rm fh}^{\rm abs}|$, implying that, as halos enter orbital motion around the filaments, the direction and magnitude of their velocity relative to the filament can change substantially along their orbits.

Strictly speaking, $v_{\rm fh}^{\rm abs}$ and $v_{\rm fh}^{\rm rel}$ are not evaluated at exactly the same time: the former corresponds to the instantaneous halo velocity in the simulation run at a given snapshot, whereas the latter is an average radial velocity of halo towards the filaments over a snapshot interval.
However, considering that the snapshot interval ($\sim 80\,\rm Myr$) is much shorter than the dynamical time for a halo to complete an orbit around the filament (several Gyr according to Figure \ref{fig:t_versus_dfh}), it should not affect the qualitative trends discussed above.

\begin{figure}
    \centering
    \includegraphics[width=0.9\linewidth]{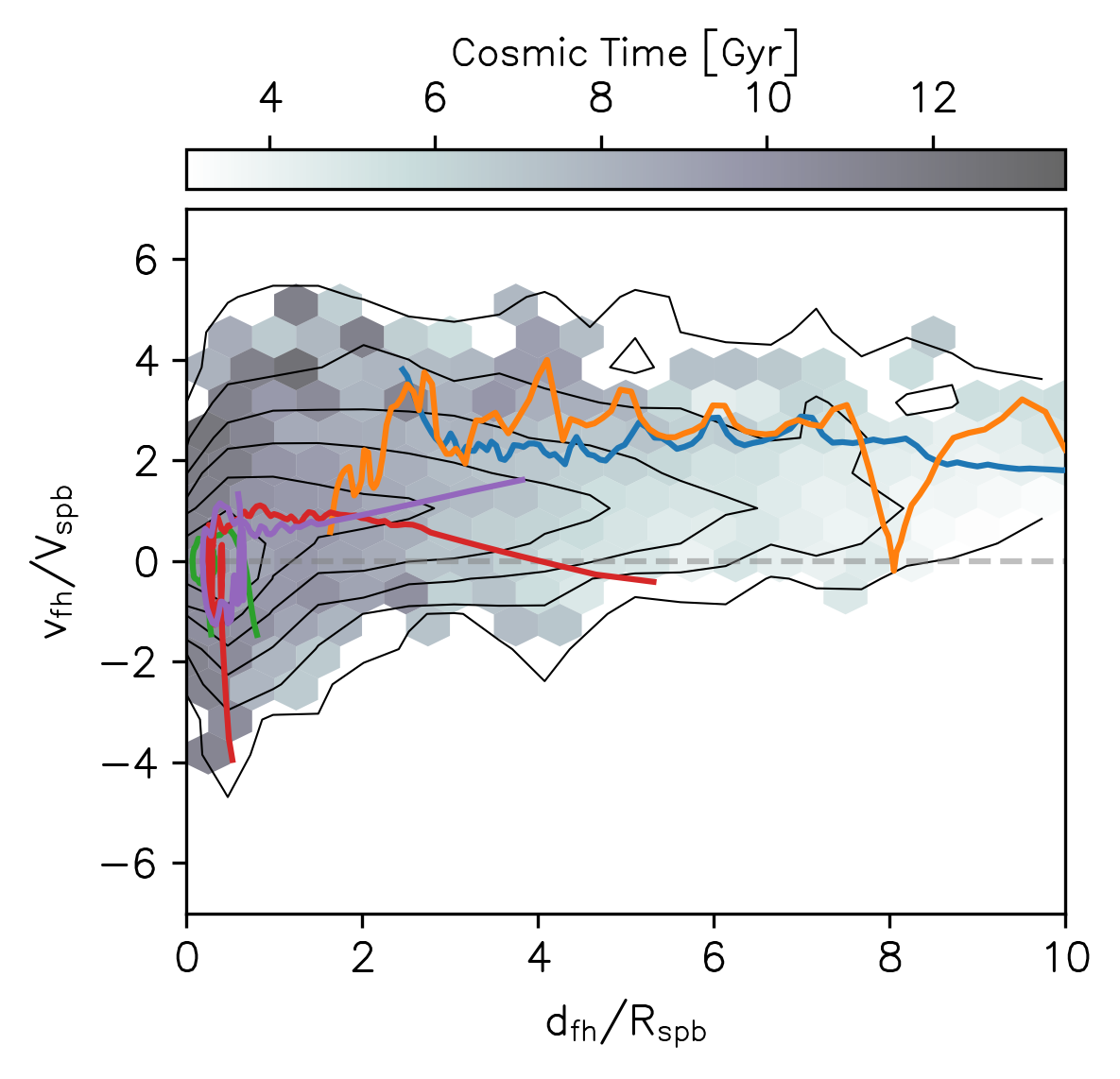}
    \caption{Stacked phase-space trajectories of 552 halos, with normalized filament-centric distances and velocities. The two-dimensional histogram is shown as black contours, while the color indicates median cosmic time within a hexagonal bin. The individual trajectories of six halos in Figure \ref{fig:t_versus_dfh} are overplotted.}
    \label{fig:phase_space_trajectory}
\end{figure}

Going beyond J22, we further normalize the phase-space coordinates to account for variations in filament radius and density across different filaments and with redshift. 
As described in Section~\ref{sec:3.2}, we define the filament splashback radius, $R_{\rm spb}$, from the logarithmic derivative of the filament density profile and adopt it as a characteristic radial scale. 
To normalize velocities in a physically motivated way, we use the mean density enclosed within $R_{\rm spb}$, $\rho_{\rm spb}$, and define a characteristic velocity scale,
\begin{equation}
    v_{\rm spb} = \sqrt{2\pi G\rho_{\rm spb}R_{\rm spb}^2},
\end{equation}
where the factor $\sqrt{2\pi}$ reflects the cylindrical geometry of filamentary structures.

Using the normalization factors, we construct normalized phase-space trajectories of halos following the approach of J22, and show the stacked distributions in Figure \ref{fig:phase_space_trajectory}.
The contours represent the number density of data points, while the hexagonal bins are colored by the mean cosmic time, allowing us to visualize the typical time ordering along the stacked phase-space pattern.
The trajectories of the six example halos in Figure \ref{fig:t_versus_dfh} are overplotted to illustrate how individual halo histories map onto the stacked distribution.

We have verified that adopting alternative density measures, such as the filament core density $\rho_{\rm core}$, leads to qualitatively similar normalized phase-space trajectories due to almost linear scaling between the two quantities.
We therefore adopt $\rho_{\rm spb}$ as a physically consistent choice tied to $R_{\rm spb}$.
Figure~\ref{fig:phase_space_trajectory} shows the resulting normalized phase-space trajectories ($d_{\rm fh}/R_{\rm spb}$, $v_{\rm fh}/v_{\rm spb}$), which display a smooth temporal progression across filaments and epochs after normalization.
The same figure without normalization is shown in the Appendix (see Figure \ref{fig:app_phase_space_trajectory}).

\subsubsection{Halo mass growth}\label{sec:4.1.2}

\begin{figure}
    \centering
    \includegraphics[width=0.99\linewidth]{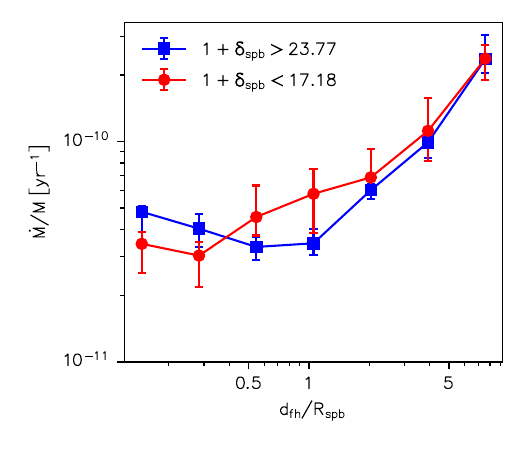}
    \caption{Specific mass accretion rates of halos as a function of normalized distances to the host filaments of two extreme present-day core densities: high-density (blue, $1+\delta_{\rm core}>23.77$) and low-density (red, $1+\delta_{\rm core}<17.18$). The solid lines represent median and uncertainties are estimated as the 16th-84th percentile range via a halo-level bootstrap with $N_{\rm bootstrap}=2000$. Mass accretion of halos around denser filaments is suppressed at $d_{\rm fh}\lesssim R_{\rm spb}$.}
    \label{fig:mass_evolution}
\end{figure}

In J22, the number density of halos associated with a host filament, $n_{\rm halo}$, emerged as a key parameter governing the mass evolution of halos in its vicinity.
Halos around the densest filaments ($n_{\rm halo}>8\,h^3\,\rm cMpc^{-3}$) exhibited substantial mass loss ($\sim 20\%$) after their first pericentric passage, which was interpreted as a signature of tidal stripping.

In this section, in order to directly assess how the filament environment affects halo mass evolution, we select halos residing in the two extreme environments: a high-density regime defined by $1+\delta_{\rm core}>23.77$ and a low-density regime defined by $1+\delta_{\rm core}<17.18$ (using the same criteria as in Figure \ref{fig:radii_evol_denprof}).
The overdensity criteria are evaluated at $z=0$.
As illustrated in the right panel of Figure \ref{fig:radii_evol_denprof}, filaments that are dense at the present epoch also tend to remain relatively dense over much of their earlier evolution, which supports the use of the $z=0$ core density as a practical selection criterion.

This selection yields 219 halos near high-density filaments and 89 halos near low-density filaments.
Figure \ref{fig:mass_evolution} shows the variation of the specific mass accretion rate, $\dot{M}/M$, as a function of the normalized distance from the filament.
In each distance bin, we first compute a representative value for each halo as the median of its measurements in that bin, and then take the median of those halo-level representative values.
The error bars indicate the 16th--84th percentile range obtained from halo-level bootstrap resampling with $N_{\rm bootstrap}=2000$.

The most noticeable difference between the two samples appears at $d_{\rm fh}/R_{\rm spb}\sim 0.6$--$2$, where halos around high-density filaments tend to exhibit lower specific mass accretion rates than those around low-density filaments, by roughly $\sim 30\%$.
This trend is consistent with the results of J22 and suggests that the suppression of halo mass growth may begin around, or slightly outside, the filament edge.
Two physical interpretations may account for this trend.
One is tidal stripping during pericentric passages through the anisotropic gravitational field of the filament, which would preferentially remove material from the outer halo.
The other is a suppression of net accretion, in the sense that dark matter flowing along the filament is simultaneously being advected toward and accreted by other nearby halos, thereby reducing the supply available to a given halo as it approaches the spine.
Because the present analysis does not allow us to distinguish robustly between these two scenarios, we refrain from attributing the observed suppression to a unique physical mechanism.
At the same time, the separation between the two samples is not statistically strong in every bin, so this result should be regarded as a moderate trend rather than a definitive distinction.

Another notable feature is that halos around low-density filaments show a sharp drop in $\dot{M}/M$ in the filament core region ($d_{\rm fh}/R_{\rm spb}\sim 0.3$--$0.4$).
This behavior may be related to the accretion histories of individual filaments or halos.
However, given the uncertainties and the smaller size of the low-density sample, it is not yet clear whether the apparent difference reflects a genuine environmental effect or is partly driven by small-number statistics.
A larger halo sample will therefore be required for a more definitive assessment.

\subsection{Differential Environment-Driven Torques }\label{sec:4.2}

\begin{figure}
    \centering
    \includegraphics[width=0.99\linewidth]{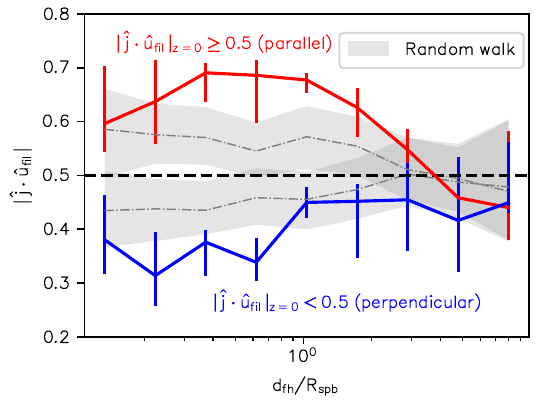}
    \caption{Halo spin–filament alignment as a function of the normalized halo–filament separation, $d_{\rm fh}/R_{\rm spb}$. Halos are subdivided at $z=0$ into a “parallel” sample (red) with $|\hat{j}\cdot\hat{u}_{\rm fil}|>0.5$ and a “perpendicular” sample (blue) with $|\hat{j}\cdot\hat{u}_{\rm fil}|<0.5$. The observed tuning-fork–like separation is compared to a null model in which halo spins evolve stochastically as an isotropic random walk on the sphere (gray region). In the random-walk model, the step angle is drawn from the empirical distribution of $\Delta\alpha$ measured from the full halo population within a $\pm 0.1\,\mathrm{dex}$ mass window around each halo. For both the simulation data and the randomized realizations, the median is shown with the 16th–84th percentile range of bootstrapped medians as the uncertainty. The alignment of each group deviates from the random sample at different locations ($\sim R_{\rm spb}$ for parallel and $\sim 0.6R_{\rm spb}$ for perpendicular).}
    \label{fig:spin_evolution}
\end{figure}

While halos move within and around filaments, the mass flows within and around the filamentary structure can torque them and modify their angular momentum.
In particular, the characteristic pattern of halo spin alignment with filament axes has been extensively studied using dark-matter-only simulations \citep[e.g.,][]{Laigle2015, Lopez2020}.
To isolate the impact of the filament environment on halo spin evolution, we exclude halos that undergo rapid mass growth between successive snapshots ($\Delta M/M>0.1$), which likely indicates recent mergers.
We are left with 148 halos whose virial masses at $z=0$ span from $7\times 10^{10}h^{-1}\,\rm M_{\odot}$ to $4.5\times 10^{12}h^{-1}\,\rm M_{\odot}$.
Although a mass-dependent “spin-flip” toward perpendicular alignments is expected at sufficiently high masses, the mass range probed here is not predicted to exhibit a strong mass dependence (i.e. $M_{\rm vir}\lesssim 10^{12.3}\,\rm M_{\odot}$; instead, halo spins are expected to be preferentially aligned parallel to nearby filaments \citep{Laigle2015}.

At each snapshot, we measure the halo spin unit vector ($\hat{j}$) and the filament-axis vector ($\hat{u}_{\rm fil}$), in order to determine whether a halo is aligned with a filament at that epoch. The alignment is quantified using the absolute value of the cosine of the angle ($\theta$) between the two vectors, $|\cos\theta| = |\hat{j}\cdot \hat{u}_{\rm fil}|$.
We define a halo spin as parallel to the filament when $|\cos\theta| > 0.5$, and as perpendicular when $|\cos\theta| < 0.5$.
Although individual halos enter the analysis at different epochs, the distribution of their initial alignment values, $|\cos\theta_{\rm init}|$, is nearly uniform between $0$ and $1$.
The spin rotation rate ($\Delta\alpha/\Delta t$) during a snapshot spans a broad range, from $10^{-3}$ to $10^{1}\,\mathrm{deg}\,\mathrm{Myr}^{-1}$.
The spins of lower-mass halos are less reliably measured because they contain fewer dark matter particles, which may contribute to apparently rapid angular changes.
This trend may be interpreted as indicating that lower-mass halos are more susceptible to changes in the angular momentum carried by externally accreted material.

We next investigate how the alignment pattern changes as halos move closer to filaments from their initial positions.
To this end, we classify the halos at the present epoch ($z=0$) according to their alignment values into a parallel group ($|\hat{j}\cdot\hat{u}_{\rm fil}|_{z=0} \geq 0.5$; 86 halos) and a perpendicular group ($|\hat{j}\cdot\hat{u}_{\rm fil}|_{z=0} < 0.5$; 62 halos).
We then trace how the median spin of these two groups has evolved in the past as a function of the normalized distance from the filament, as shown in Figure \ref{fig:spin_evolution}.
In this plot, the parallel and perpendicular groups are shown by the red and blue curves, respectively.

Overall, we find that the bootstrap results for the two groups converge toward the median value of $\sim 0.5$ at around $d_{\rm fh}/R_{\rm spb} \sim 1$.
As a result, the dependence on distance exhibits a ``tuning-fork''-like pattern.
This shape is not, by itself, indicative of any peculiar physical property, but rather arises naturally when one traces backward in time a sample that has been subsampled according to its present-day values.
For example, such a pattern would also be expected even if halo spins evolved randomly from the same initial distribution.
Moreover, regardless of whether the spin evolution is random or non-random, the location of the bifurcation is determined by how long the spin vectors retain memory of their past states, or equivalently, by how rapidly the spin directions rotate.
It is therefore important to examine whether the tuning-fork pattern traced by the red and blue curves in this figure differs from that expected in a model in which the spin evolution is randomized while preserving the same spin rotation rate.

To test this directly, we performed an experiment in which halo spins evolve from their initial orientations with random azimuthal changes.
As noted above, the key issue is how to incorporate the angular change $\Delta\alpha$ measured from the actual spin evolution into the random-evolution model.
This is because the properties of the random-evolution model can vary substantially depending on how $\Delta\alpha$ is treated.
In practice, the time intervals between snapshots are not uniform, and the spin variation rate may also depend on halo mass.
We therefore used the halo-mass-dependent distribution of $\Delta\alpha/\Delta t$ derived from the actual spin evolution. 
Specifically, for each halo, we randomly drew a value from the distribution of actual spin variation rates of halos whose masses agree within $\pm 0.1\,\mathrm{dex}$, and used this value together with the time interval $\Delta t$ to the next snapshot to determine the angular change $\Delta\alpha$.
The spin vector was then evolved such that the magnitude of the rotation angle was fixed to $\Delta\alpha$, while the azimuthal direction of that change was assigned randomly, thereby implementing a random evolution independent of the filament direction.
We note that, although the filamentary environment may in principle affect $\Delta\alpha/\Delta t$ itself, this possibility is neglected in the present model, where we consider only the influence of the filament on the \textit{direction} of spin evolution.

The resulting expectation from this null test is shown as the gray band in Figure \ref{fig:spin_evolution}.
Although a tuning-fork--like separation naturally arises under purely stochastic evolution, the observed separation in the simulation exceeds the null expectation, with the two tracks lying outside each other's $\sim 1\sigma$ bands over a range of filament-centric distances.
This result supports an environmentally driven picture of halo spin reorientation and complements previous studies of spin--filament alignments \citep[e.g.][]{Codis2015, Laigle2015} by explicitly following individual dark matter halos in an evolving filament frame.
Interestingly, the departures from the null model occur at different halo–filament separations for the two subsamples: the “parallel” population shows a clear excess relative to the null expectation already at $d_{\rm fh}\simeq 0.4$–$1.0\,R_{\rm spb}$, whereas the “perpendicular” population departs mainly at $d_{\rm fh}\simeq 0.2$–$0.6\,R_{\rm spb}$.
A plausible interpretation is that the dominant flow of the surrounding matter changes with distance to the filament spine.
Near the filament boundary, multistreaming can generate coherent transverse (azimuthal) swirling motions, with vorticity preferentially aligned with the filament axis, which can torque halo spins and enhance a parallel alignment signal.
Deeper in the filament interior, increasingly strong axial transport toward nodes may provide an additional channel for spin reorientation, potentially contributing to the delayed emergence of the perpendicular deviation at smaller $d_{\rm fh}$.

\section{Summary} \label{sec:5}
We study the coevolution of dark matter halos and large-scale cosmic filaments in a time-dependent, comoving reference frame defined by evolving filament progenitors.
To enable such analyses, we develop and apply a methodology to identify and track the progenitors of individual filaments across cosmic time. Using this framework, we perform a feasibility study to assess filament evolution itself and to revisit halo dynamics, mass growth, and spin evolution in filamentary environments.

We apply our method to the dark-matter-only $N$-Cluster Run simulation, selecting three cluster-centered subvolumes and extracting filamentary structures at all snapshots from $z=3$ to $z=0$ using \texttt{DisPerSE}.
Our approach explicitly links each present-day filament to its progenitors by maximizing a spatial-similarity score based on the three-dimensional overlap of filament segments. When necessary, we iteratively relax the \texttt{DisPerSE} persistence threshold to recover less prominent progenitors, enabling us to trace a substantial subset of filaments back to $z>1$.

Using the identified filament progenitors, we found:
\begin{enumerate}
    \item Filament evolution and drift: We measure filament peculiar drift velocities using the motions of critical points, finding typical values of a few hundred $\rm km \,s^{-1}$, comparable to halo peculiar velocities. This highlights the importance of accounting for filament advection when interpreting filament-centric halo dynamics.
    \item Time-resolved filament structure: By reconstructing radial density profiles from dark matter particles, we identify a mild expansion of the filament splashback radius from $\sim0.7\,\rm cMpc$ at $z\gtrsim1$ to $\sim 1.2\,\rm cMpc$ at $z=0$. The evolution of core densities follows a divergent pattern: high-density filaments become increasingly dense through further accretion, whereas low-density filaments tend to become more diffuse. We show that while average filament growth broadly follows previously reported stacked trends, individual filaments exhibit substantial diversity tied to their late-time accretion histories.
    \item Halo orbital dynamics in an evolving filament frame: Revisiting halo trajectories using progenitor-based filament distances, we find that assuming filaments fixed at $z=0$ systematically biases inferred halo-filament separations and orbital phases, particularly at high redshift. Filament drift leaves a clear imprint on halo velocities at distances $\gtrsim 1-2 R_{\rm spb}$. Adopting the evolving filament frame further enables more coherent stacking of halo phase-space trajectories across different filaments and epochs, reducing artificial smearing introduced by filament drift and variations in filament size and density.
    \item Halo mass growth regulation: We split halos into two groups residing near high- and low-density filaments, defined by the present-day filament overdensity, and examined how the specific mass accretion rate ($\dot{M}/M$) varies with normalized distance from the filament. Halos near high-density filaments show a systematically lower $\dot{M}/M$ already at $d_{\rm fh}/R_{\rm spb}\sim 0.6-2$, suggesting that the suppression of mass growth sets in at the filament outskirts.
    \item Halo spin--filament alignment: We track the evolution of halo spin--filament alignment, quantified by $|\hat{j}\cdot\hat{u}_{\rm fil}|$, for a merger-suppressed sample of 148 halos. We find that halo spin alignment exhibits a tuning-fork--like bifurcation as halos approach filaments when halos are subdivided by their present-day alignment. A mass-conditioned random-walk null model naturally reproduces part of this separation due to selection bias. However, the observed bifurcation significantly exceeds the null expectation over characteristic filament-centric distances, indicating that filament-driven torques contribute to halo spin reorientation during infall.
\end{enumerate}

\section{Acknowledgements}
This work is partially supported by the grant GALBAR ANR-25-CE31-4684 and IDF-DIM-ORIGINES-2023-4-11 and from the CNRS through the MITI interdisciplinary programs.
This work is  partially supported by the National Research Foundation of Korea(NRF) grant funded by the Korea government (MSIT) (No. RS-2024-00455106).
This research was supported by Global - Learning \& Academic research Institution for Master's·PhD students, and Postdocs(G-LAMP) Program of the National Research Foundation of Korea(NRF) grant funded by the Ministry of Education(No. RS-2025-25442707).
EC acknowledges support from the National Research Foundation of Korea (NRF-RS-2025-00515276).

\section{Data Availability}
The data underlying this article will be shared on reasonable request to the corresponding author.

\begin{appendix}
\section{An algorithm to trace the filament progenitors} \label{sec:ap:A}

\subsection{Spatial Similarities}\label{sec:ap:A1}
Assuming that the positions of the large-scale structures do not dramatically change during the time separation of two consecutive snapshots ($\sim 64$--$110\,\rm Myr$), tracing the progenitors of a filament can be approached as identifying the most spatially similar structures across time.
The similarity is quantified by score $\mathcal{S}$, based on the concept of Bhattacharyya coefficient \citep{Bhattacharyya1943}, which measures the similarity between two probabilistic distributions $p(x)$ and $q(x)$:
\begin{equation} \label{eq:BC}
    BC(p,q) = \sum_x \sqrt{p(x)q(x)}\,.
\end{equation}
Basically, it measures the joint probability of two independent probability distributions.
Similarly, we can define the `spatial' similarity between two filaments (labeled as 1 and 2, respectively) by multiplying the two normalized histograms of their positions along e.g., $x$-axis,:

\begin{equation} \label{eq:Sx}
    S_x = \sum_{i=0}^{n_{\rm bin}} \sqrt{\left(\frac{n_{1,i}}{N_1}\right)\left(\frac{n_{2,i}}{N_2}\right)}\,,
\end{equation}
where $n_{\rm bin}$ is the number of histogram bins along $x$-axis, $N_1$ and $N_2$ are the number of segments of the two filaments, and $n_{1,i}$ is the number of segments of the first filament within $i$-th bin.
The sum is taken over a set of bins defined for the entire range that includes both filaments.
The total score $\mathcal{S}$ is calculated as a product of scores along the three axes:
\begin{equation} \label{eq:Score}
    \mathcal{S}=(\mathcal{S}_x\mathcal{S}_y\mathcal{S}_z)^{1/3}.
\end{equation}
By construction, $\mathcal{S}$ is non-zero only if the spatial extents of the two filaments overlap along \emph{all three} axes (i.e., $S_x,S_y,S_z\neq 0$); otherwise, $\mathcal{S}=0$ if the overlap vanishes along any one axis.
The score becomes one if the two filaments are perfectly the same.
A progenitor of a given filament at a snapshot is then ideally chosen as the one with the highest score among progenitor candidates in an earlier snapshot.

When calculating $\mathcal{S}$, the number of bins must be chosen carefully, taking into account the number of segments in the two filaments being compared.
If the number of bins is too large, some bins may remain empty, containing no filament segments, simply because the filaments are not densely sampled.
Conversely, if the number of bins is too small, most bins will contain filament segments, causing $\mathcal{S}$ to appear high even when the two filaments are not similar at all.
We optimized the number of segments and, consequently, the number of bins by analyzing how $\mathcal{S}$ changes for different numbers of segments and bins.
These two parameters are controlled by $\Delta s$ (the segment length) and $\Delta x$ (the bin width).
Although this depends significantly on the shape and length of filaments, we were able to quantify the similarity between two given filaments reasonably when $\Delta s$ is small enough and $\Delta x$ is larger than $\Delta s$. 
We set $\Delta s=1/40 \,h^{-1}\,{\rm cMpc}$  and $\Delta x=2\Delta s$.

\subsection{Looping over Snapshots}\label{sec:ap:A2}
Using the scoring method described in Section \ref{sec:ap:A1}, the progenitors can be traced by subsequently choosing a candidate with the highest spatial similarity.
However, one might face cases where the progenitor candidate with the highest score still does not match well with the descendant.
It could imply that the filament does not have a well-identified progenitor; however, it may also occur if the progenitor's persistence is lower than the given {\tt -cut} value, as the density contrast tends to decrease to higher redshifts.
In this case, we may need to adjust the persistence threshold when none of the progenitor candidates meet the minimum similarity requirement, as quantified by the score threshold, $\mathcal{S_{\rm thr}}$.

That is, even when we have found a progenitor with the highest non-zero score, we will not choose it if $\mathcal{S}<\mathcal{S}_{\rm thr}$.
Instead, we re-extract filaments with a lower persistence threshold ({\tt cut}=$c-\Delta c$) and assess the similarity of the new progenitor candidates with the target descendant filament.
The persistence threshold gets lower until a proper progenitor with $\mathcal{S} \geq \mathcal{S}_{\rm thr}$ is found, and the tracking is finished when the persistence becomes zero.
The final snapshot traced back for the progenitor could be different from one filament to another: some filaments end within several snapshots, while others can be traced back to $z=3$.

We tested several values of $\mathcal{S}_{\rm thr}$ and ultimately adopted a fixed value of 0.8 for all redshifts and filaments.
The initial {\tt -cut} parameter is set to 1.0, such that we begin with filament candidates whose node--saddle density contrast corresponds to a $1\sigma$ threshold.
During the procedure, we decrease this threshold by $\Delta c=0.1$ whenever a suitable progenitor is not found.

We emphasize that the choices of the parameters ($\Delta s$, $\Delta x$, $\mathcal{S}_{\rm thr}$, and $\Delta c$) are entirely empirical: different settings can lead to different outcomes in the individual-filament levels, reflecting the diverse geometries.
After exploring a wide range of values, we find that the adopted configuration yields robust and satisfactory tracing results overall.

Finally, our algorithm traces only the \emph{most prominent} progenitor for each filament and does not attempt to construct a full filament merger tree, which is left to future work.

\begin{figure}
    \includegraphics[width=0.99\linewidth]{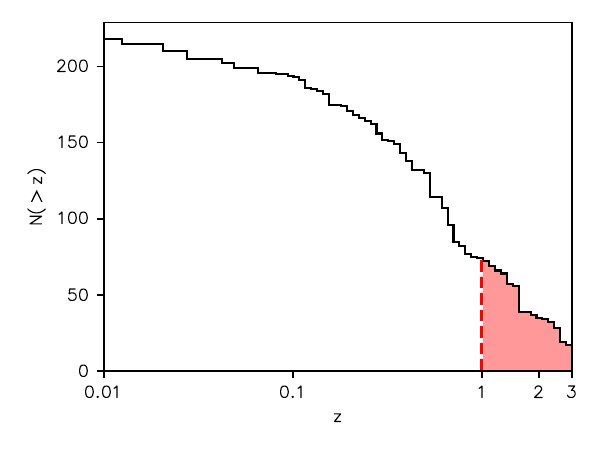}
    \caption{The number of filaments at $z=0$ whose progenitors have been traced back beyond the redshift $z$ indicated in the x-axis. The red vertical line indicates $z=1$. $N(z>1)=73$ filaments have their initial progenitor at $z>1$.}
    \label{fig:num_of_fil_progs}
\end{figure}

\section{Supplementary figures}
\subsection{Filament evolution}
\begin{figure*}
    \centering
    \includegraphics[width=0.99\linewidth]{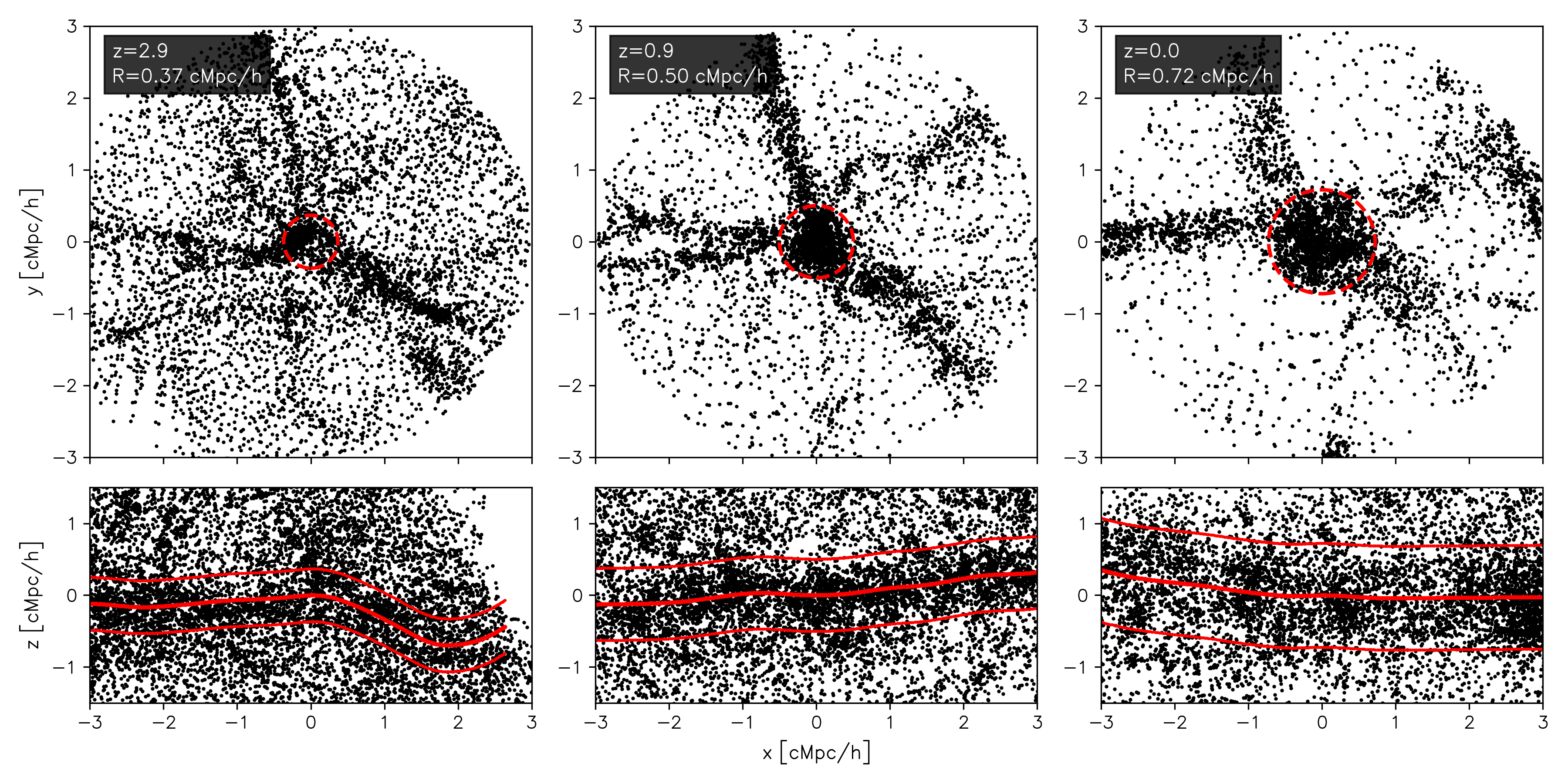}
    \caption{The upper row shows the distribution of dark matter particles (black scatters) within a $3\,{\rm cMpc}/h$-width slab in a plane perpendicular to the filament. The bottom row presents the longitudinal view of the filament, where the red thin line shows the filament spine. From left to right, the panels correspond to redshifts $z=2.9$, $0.9$, and $0.0$. The red dashed lines mark the filament splashback radius in both the upper and lower panels, identified at each redshift as the radius where $\gamma$ attains its minimum, corresponding to the steepest logarithmic slope of the transverse density profile.}
    \label{fig:cross_section}
\end{figure*}

Figure~\ref{fig:cross_section} illustrates the dark matter particle distribution around a representative filament at different redshifts.
The upper panels show cross-sections perpendicular to the filament spine, while the lower panels show the corresponding longitudinal views.
The filament position and radius are indicated by red lines.

\subsection{Unnormalized phase-space trajectories of halos around filaments}
\begin{figure*}
    \centering
    \includegraphics[width=0.9\linewidth]{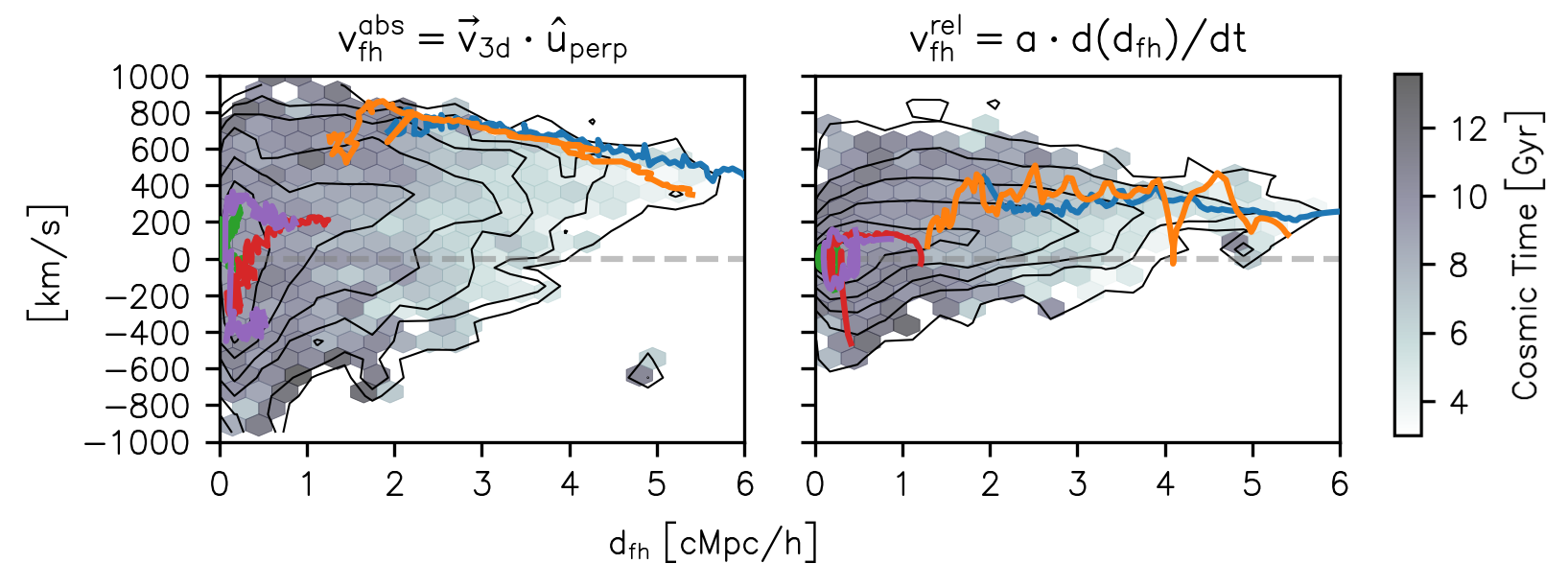}
    \caption{Stacked phase-space trajectories of 552 halos, without normalization. In the left panel, $y$-axis is the absolute velocity component of halo 3D velocity projected to $\hat{u}_{\rm perp}$, while in the right panel, the relative velocity of halos with respect to filaments is used instead. Compared to Figure \ref{fig:phase_space_trajectory}, the color gradient is noisier.}
    \label{fig:app_phase_space_trajectory}
\end{figure*}

Figure~\ref{fig:app_phase_space_trajectory} shows the stacked halo phase-space trajectories before applying the filament-dependent normalization by $R_{\rm spb}$ and $v_{\rm spb}$.
Compared with the normalized version in Figure~\ref{fig:phase_space_trajectory}, the unnormalized distribution exhibits a noisier time ordering, reflecting the diversity of filament sizes, densities and drift motions.
\clearpage
\end{appendix}



\bibliographystyle{mnras}
\bibliography{sample631} 


\appendix
\label{sec:appendix}

\bsp	
\label{lastpage}

\end{document}